\documentclass[letterpaper,twocolumn,10pt]{article}
\usepackage{qwin}

\usepackage{tikz}
\usepackage{amsmath}

\usepackage{caption}
\usepackage{graphicx}
\usepackage{float} 
\usepackage{subfigure}

\usepackage{setspace}
\usepackage{url}
\usepackage{cleveref}
\crefname{section}{§}{§§}
\Crefname{section}{§}{§§}
\usepackage[ruled,linesnumbered]{algorithm2e} 
\usepackage{multirow}
\usepackage{authblk}

\begin{document}

\date{}

\title{\Large \bf QWin: Enforcing Tail Latency SLO at Shared Storage Backend}

\author{
Liuying Ma, Zhenqing Liu, Jin Xiong, Dejun Jiang
}
\affil{
Institute of Computing Technology, Chinese Academy of Sciences, Beijing
}
\affil{
maliuying, liuzhenqing, xiongjin, jiangdejun@ict.ac.cn
}

\maketitle

\gdef\LSA{LSW}
\gdef\BOA{BOW}

\begin{abstract}

Consolidating latency-critical (LC) and best-effort (BE) tenants at storage backend helps to increase resources utilization. 
Even if tenants use dedicated queues and threads to achieve performance isolation, threads are still contend for CPU cores. 
Therefore, we argue that it is necessary to partition cores between LC and BE tenants, and meanwhile each core is dedicated to run a thread. 
Expect for frequently changing bursty load, fluctuated service time at storage backend also drastically changes the need of cores.
In order to guarantee tail latency service level objectives (SLOs), the abrupt changing need of cores must be satisfied immediately. Otherwise, tail latency SLO violation happens.
Unfortunately, partitioning-based approaches lack the ability to react the changing need of cores, resulting in extreme spikes in latency and SLO violation happens. 
In this paper, we present QWin, a tail latency SLO aware core allocation to enforce tail latency SLO at shared storage backend.
QWin consists of an SLO-to-core calculation model that accurately calculates the number of cores combining with definitive runtime load determined by a flexible request-based window, and an autonomous core allocation that adjusts cores at adaptive frequency by dynamically changing core policies.
When consolidating multiple LC and BE tenants, QWin outperforms the-state-of-the-art approaches in guaranteeing tail latency SLO for LC tenants and meanwhile increasing bandwidth of BE tenants by up to 31x. 
\end{abstract}

%-------------------------------------------------------------------------------
%\section{Introduction}
%-------------------------------------------------------------------------------

\section{Introduction}
Distributed storage systems are widely deployed as the storage infrastructure in clouds. 
Typically, they provide virtual disks attached to virtual machines or containers in a multi-tenant way ~\cite{pslo16,smite14,cake12,reflex17,pm14,sncm16,dishangyun}.
Different tenants have distinguished performance requirements for virtual disks. 
For example, latency-critical (LC) tenants (e.g. Web-server/OLTP) require low latency guarantees, and meanwhile best-effort (BE) tenants expect sustainable bandwidth. 
The performance requirements are usually defined as Service Level Objectives (SLOs). 
In order to guarantee user's experience, tail latency SLO at 99th and 99.9th percentiles is becoming the main concern of today's cloud providers~\cite{dean13,dynamo07,tales14,cpi2}. 

Storage backend (e.g. Ceph's OSD~\cite{ceph06}, HDFS's DateNode~\cite{hdfs10}) always plays an important role in distributed storage system. 
Storage backend receives IO requests from upper-level applications, and then processes them by accessing underlying storage devices (e.g. HDDs or SSDs).
Storage backend is usually constructed using the stage event-driven architecture (SEDA)~\cite{seda01}. In SEDA architecture, each stage consists of a queue and a thread pool which is responsible for request processing. A request is always queued until a CPU core is available and an idle thread is scheduled to run. As a result, the latency of a request is affected by CPU core allocation as well as thread scheduling.

When serving LC and BE requests simultaneously, existing works try to guarantee tail latency SLO by priority-based scheduling~\cite{cake12,reflex17,pm14,pslo16,sncm16}. 
Generally, LC and BE requests are usually partitioned in respective queues, but threads are still shared. 
A few consecutive BE requests (e.g. 64KB or larger) can block subsequent LC requests (e.g. 4KB or smaller). The latencies of LC requests increase, which can cause tail latency violation.
Even if threads are also partitioned among LC and BE tenants, they always contend for CPU cores.
When threads happen to be scheduled to process BE requests, CPU cores may be all occupied by these threads. Consequently, threads processing LC requests cannot be scheduled since no CPU cores are available.
This seriously affects the latencies of LC requests, and tail latency SLO violation can also happen.

The conventional wisdom is that cores are statically partitioned between LC and BE tenants. And each core is dedicated to run a thread.
However, many tenants experience bursty request patterns or phased behavior, drastically changing the amount of cores they need. 
Moreover, the underlying storage device is shared. The service time (i.e. I/O latency) fluctuates greatly, specially SSDs. Large fluctuated service time makes backlog of requests, and more cores are needed to process these requests quickly.
Therefore, cores must be dynamically allocated to respond to these situations. 
In such cases, two issues must be addressed. 

First, it is necessary to accurately calculate the number of cores and adjust cores to respond to the changing need. Recent works~\cite{cake12,heracles15,parties19} adopt a incremental core adjustment approach based on historical information at fixed intervals, which cannot immediately derive the accurate cores when the need of cores is changed.

Second, unlike core allocation for in-memory applications~\cite{perfiso18,arachne18,shenango19,heracles15,parties19,caladan20} which only considers the bursty load, core allocation on storage backend must take the fluctuated service time (which has significant impact on latency) into account.

In order to guarantee tail latency SLO for LC tenants meanwhile maximizing bandwidth of BE tenants, we present QWin, a tail latency SLO aware core allocation for shared storage backend.
QWin proposes two key ideas. 

First, QWin builds an SLO-to-core calculation model which can accurately calculates the number of cores for LC tenants.
Except for target SLO, this model also needs definitive runtime load as a parameter. To this end, we propose a flexible request-based window to accurately quantify the runtime load. 
To our best knowledge, \textit{QWin is the first approach to accurately translate target SLO into the number of cores}.

Second, for fast detection and targeted reactions to bursty load and fluctuated service time, QWin provides three core policies for adjusting cores. Core policy determines the frequency to check if the need of cores is changed and how many cores are needed.
QWin dynamically changes core policies at different stages to adapt to the changing need of cores. 
Core adjustment is autonomous, and no dedicated core is required to collect information and adjust cores.

We implement QWin algorithm in the storage backend (OSD~\cite{rados07}) of widely-used open-source distributed storage system Ceph~\cite{ceph06}. Using both Fio~\cite{fio} and workloads from Filebench~\cite{filebench} to simulate LC and BE tenants, we evaluate QWin, and compare it with the-state-of-the-art approaches. The experimental results show that QWin is highly general, supporting multiple workloads. Further, QWin outperforms these approaches, with simultaneously enforcing diverse target SLOs for multiple LC tenants and enhancing bandwidth of BE tenants, e.g. the bandwidth of BE tenants is increased by up to 31x without target SLO violations.

\section{Background and Motivation}

\subsection{Basis of Storage Backend Core Allocation} 
\begin{figure}[h]
\setlength{\abovecaptionskip}{0.1cm}
\centerline{\includegraphics[width=0.5\textwidth,trim=0 400 0 145,clip]{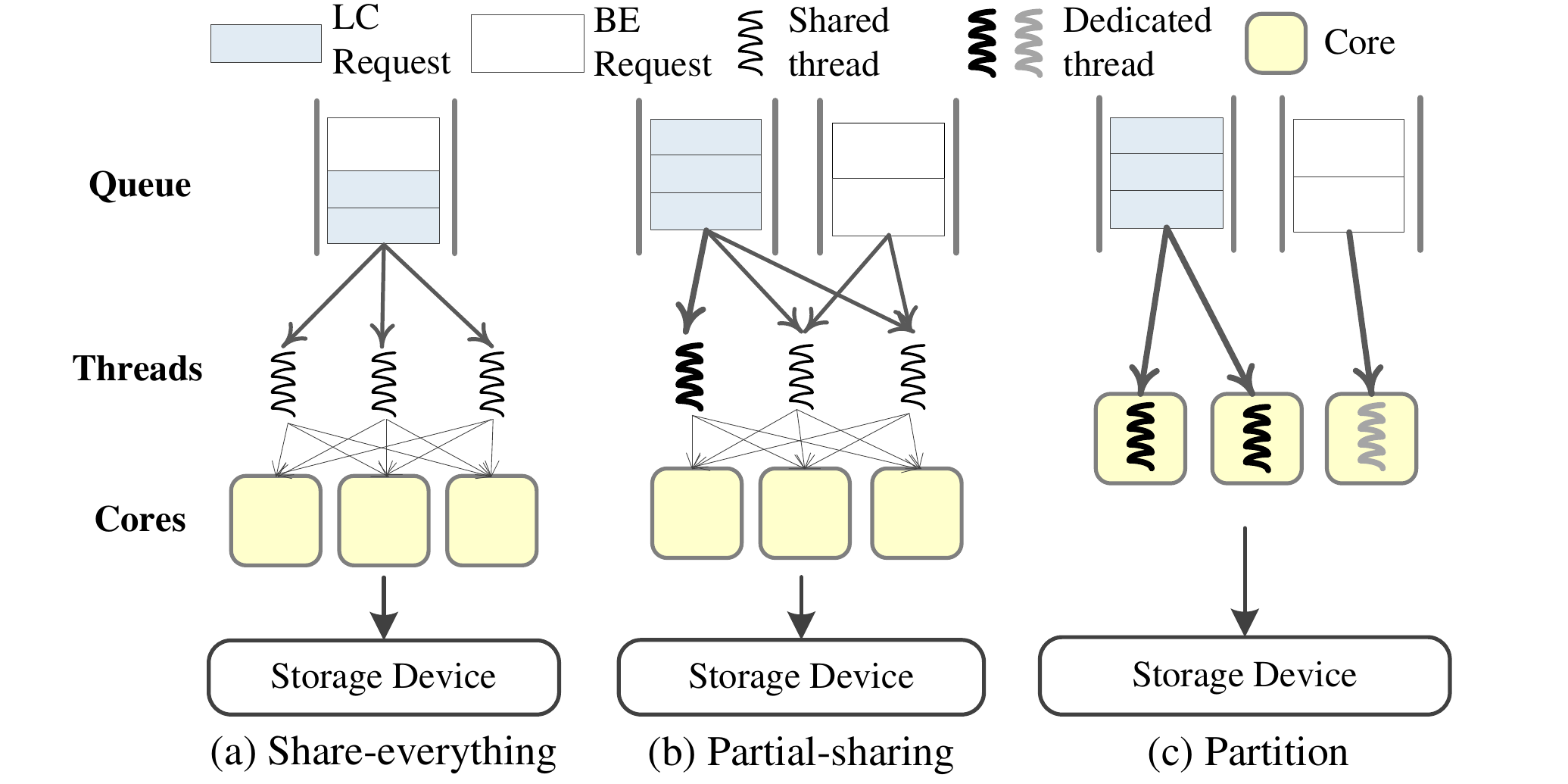}}
\caption{Resource sharing model in storage backend.}
\label{qtc}
\end{figure}

Within a storage backend of distributed storage system (e.g. Ceph's OSD), the request processing comprises three stages: 1) a request is received from network and enters into queue waiting for being assigned a thread; 2) once a thread is available, the request is dequeued and processed by local storage engine (e.g. Ceph's Bluestore~\cite{cephsosp19}); 3) the response is sent back. 
At runtime, a thread runs on a core and consumes requests from queues. When serving both LC and BE tenants, resources can be shared in three ways as shown in Figure~\ref{qtc}. 

In \textit{Share-everything} model (Figure~\ref{qtc}(a)), requests of LC tenants are not distinguished from BE tenants, and threads are not dedicated to any queue nor any CPU core. In such case, requests from different tenants compete queues and threads as well as CPU cores. 
In \textit{Partial-sharing} model (Figure~\ref{qtc}(b)), LC and BE requests enter into dedicated queues may have dedicated threads. 
However, competition cannot be avoided since threads can still run on any CPU core.
These two models suffer from resources (e.g. cores and threads) sharing between LC and BE tenants. This results in inter-tenant interference, making tail latency unpredictable.
In \textit{Partition} model (Figure~\ref{qtc}(c)), all resources are partitioned between tenants. A CPU core is dedicated to run one thread to avoid switching. Each tenant type has its own queue and cores (threads). Note that a tenant type represents a group of tenants with same performance requirement.
Although static partition can reduce interference, it sacrifices resource utilization because each tenant must be provisioned enough cores to accommodate peak load. 
Previous works~\cite{arachne18,shenango19,perfiso18,caladan20,heracles15,parties19,cake12,reflex17} dynamically adjust core allocation among tenants to maximize resource utilization. 
Unfortunately, all of these approaches do not precisely calculate core requirements according to target SLO, and/or do not target requests accessing the underlying storage devices.

\subsection{Limitations of Existing Approaches}
\begin{figure}[ht] 
  \centering
  \setlength{\abovecaptionskip}{-0.1cm}
  \subfigbottomskip=-1pt
   \subfigure{
    \includegraphics[width=0.48\textwidth,trim=2 35 78 95,clip]{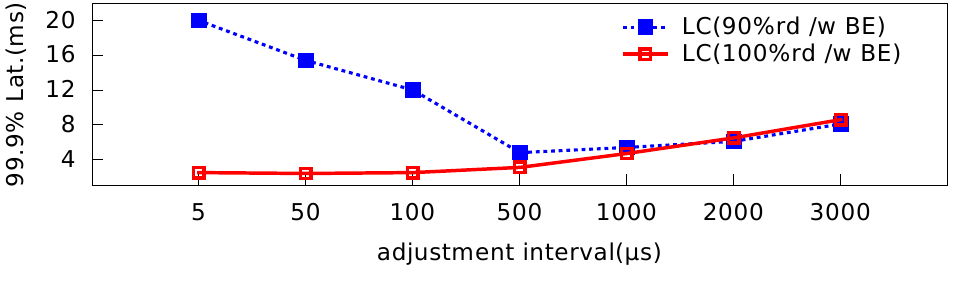}}
  \caption{Tail latency changes with different adjustment intervals under two scenarios using Shenango's core allocation algorithm: 1) consolidating a 100\% read LC tenant (A in Table~\ref{workload}) and a BE tenant; 2) consolidating a 90\% read LC tenant (C in Table~\ref{workload}) and a BE tenant. The BE tenant issues workload E in Table~\ref{workload}.}
  \label{shenango} 
\end{figure}

\textbf{SLO-unaware core allocation.}
Existing works do not consider target SLO when allocating cores among multiple tenants~\cite{arachne18,shenango19,perfiso18,caladan20}.
These systems all target to minimize tail latency for LC tenants. They use metrics that are independent of target SLO such as core utilization, queuing delay or runtime load to reallocate cores at fixed adjustment intervals. 
However, there are two main issues. 
First, tail latencies of LC tenants are suffered by different adjustment interval. 
The interval needs to be manually tuned for different tenants to minimize tail latency.
As shown in Figure~\ref{shenango}, for read-only LC tenant (100\% read), the short the interval, the lower the tail latency; but for read-heavy LC tenant (90\% read), tail latency is high when interval is short.
This is because such approaches allocate cores without considering the fluctuated service time from the underlying storage devices.
Thus, when consolidating multiple tenants, it is hard to set a proper adjustment interval for all LC tenants to minimize tail latencies, respectively.
Second, core allocation that only target to minimize tail latency results in low utilization. 
Actually, if LC tenants have looser tail latency requirement, it means that requests can be queued for a certain while instead of being processed immediately. Consequently, the number of cores for satisfying a looser tail latency is less than that for minimizing tail latency. 
However, such SLO-unaware core allocation makes that LC tenants occupy more cores than the actual need. This results in lower bandwidth of BE tenants.

\textbf{SLO-aware core allocation.} 
Previous works dynamically allocate cores among multiple tenants according to target SLO~\cite{heracles15,parties19}.
However, these solutions do not calculate the number of cores according to target SLO. They adopt trial-and-error approach to adjust cores.
Meanwhile, each core adjustment is incremental, and converging to the right allocations takes dozens of seconds, which cannot satisfy millisecond timescales tail latency requirement. 

\begin{figure}[ht] 
  \centering
  \setlength{\abovecaptionskip}{-0.1cm}
  \subfigbottomskip=-1pt
   \subfigure{
    \includegraphics[width=0.48\textwidth,trim=2 65 78 92,clip]{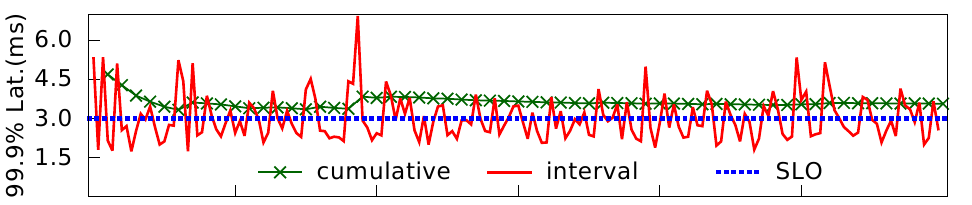}}
  \subfigure{
    \includegraphics[width=0.48\textwidth,trim=2 35 78 90,clip]{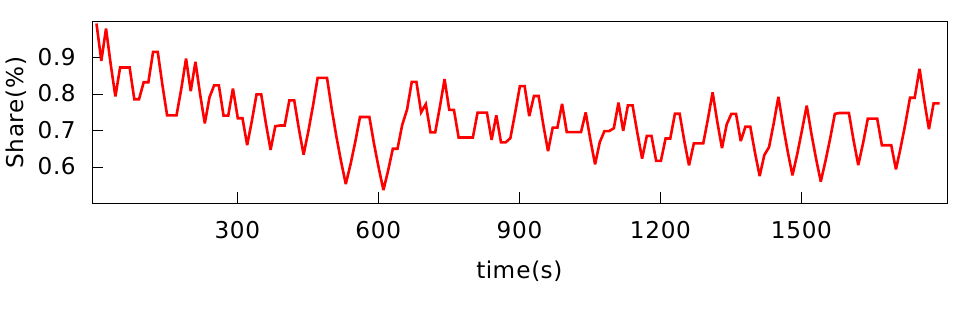}}
  \caption{Performance of Cake when consolidating an LC tenant (A in Table~\ref{workload}) and a BE tenant (E in Table~\ref{workload}). The upper shows the cumulative 99.9th latency, the 99.9th latency in every $10s$ interval and the target SLO ($3ms$) of the LC tenant. The lower shows the proportional share of cores used for the LC tenant.}
  \label{cake} 
\end{figure}

\textbf{SLO-aware core sharing approaches.} 
Some existing works adopt a sharing model where cores or threads are shared to dynamically schedule LC and BE requests according to target SLO~\cite{cake12,reflex17,pm14,pslo16,sncm16}.
As shown in Figure~\ref{cake}, Cake~\cite{cake12} adjusts proportional shares and reservations of cores (reservations are set only if shares are found to be insufficient) for an LC tenant based on the historical statistics (tail latency in the previous interval). There is no reserved core for the LC tenant in this experiment.
Unfortunately, in most cases, as shown in the upper, tail latencies of two consecutive intervals are significantly different, leading to improper core adjustment. This results in SLO violation in subsequent interval.
Although approaches such as rate limiting and priority are combined to schedule LC and BE requests, 99.9th and higher percentile tail latencies suffer since they are more sensitive to the sharing resources. 
As a result, target SLO is hard to be satisfied, or resource utilization is low due to over-provisioning resources for enforcing target SLOs.

\subsection{Challenges}
\label{challenges}
In order to enforce tail latency SLO for LC tenants meanwhile maximizing bandwidth for BE tenants, we had to overcome three key challenges: 

\textbf{Translate target SLOs into core requirements.} 
Target SLO actually reflects core requirement. Meanwhile, core requirement is also affected by other factors, such as the runtime load.
Previous works tried to enforce target SLO either by static over-provision~\cite{op14}, or dynamically adjusting cores by using historical tail latencies~\cite{cake12,heracles15,parties19}. 
However, these works cannot decide the accurate core requirement.
As a result, either SLO cannot be satisfied or resources are wasted.
It is necessary to decide which factors are used to accurately calculate core requirement. Meanwhile, these factors need to be precisely quantified at runtime with low overhead.

\textbf{Runtime load quantification.} 
Request rates issued to storage backends are not fixed, and core allocation should be changed with the load. 
Existing systems~\cite{arachne18,shenango19,heracles15,parties19,cake12} adjust cores at fixed intervals (e.g. $10s$ for Cake~\cite{cake12} and $5~\mu s$ for Shenango~\cite{shenango19}). 
Nevertheless, the runtime load is always changing within an interval due to the incoming requests and the completed requests.
No matter how short/long the interval is, the load cannot be precisely quantified. Therefore, it is necessary to design precise load quantification approach.

\textbf{Fast and precise reactions to bursty loads and fluctuated service time.}
Workload Bursts are very common with bursting interval at microsecond timescales~\cite{shenango19,perf18,perfiso18,caladan20}. 
Meanwhile, for typical NVMe Flash devices used in storage backend, service time (usually hundreds of microseconds) fluctuations are normal~\cite{st-f2,st-f3,st-f4,st-f5,reflex17}. 
Both bursty load and fluctuated service time create a backlog of requests, resulting in long queuing delay.
Prior works adjust cores incrementally for all tenants at fine time scales to handle bursty load without I/O accessing~\cite{shenango19,caladan20}. 
However, incremental core adjustment cannot immediately satisfy the bursts of core usage under bursty load or with fluctuated device performance. Therefore, it is necessary to design fast and precise core adjustment approach to handle bursty loads and fluctuated service time.

\section{Design}

\subsection{QWin Overview}
\begin{figure}[htbp]
\setlength{\abovecaptionskip}{0.1cm}
\centerline{\includegraphics[width=0.5\textwidth,trim=65 500 10 82,clip]{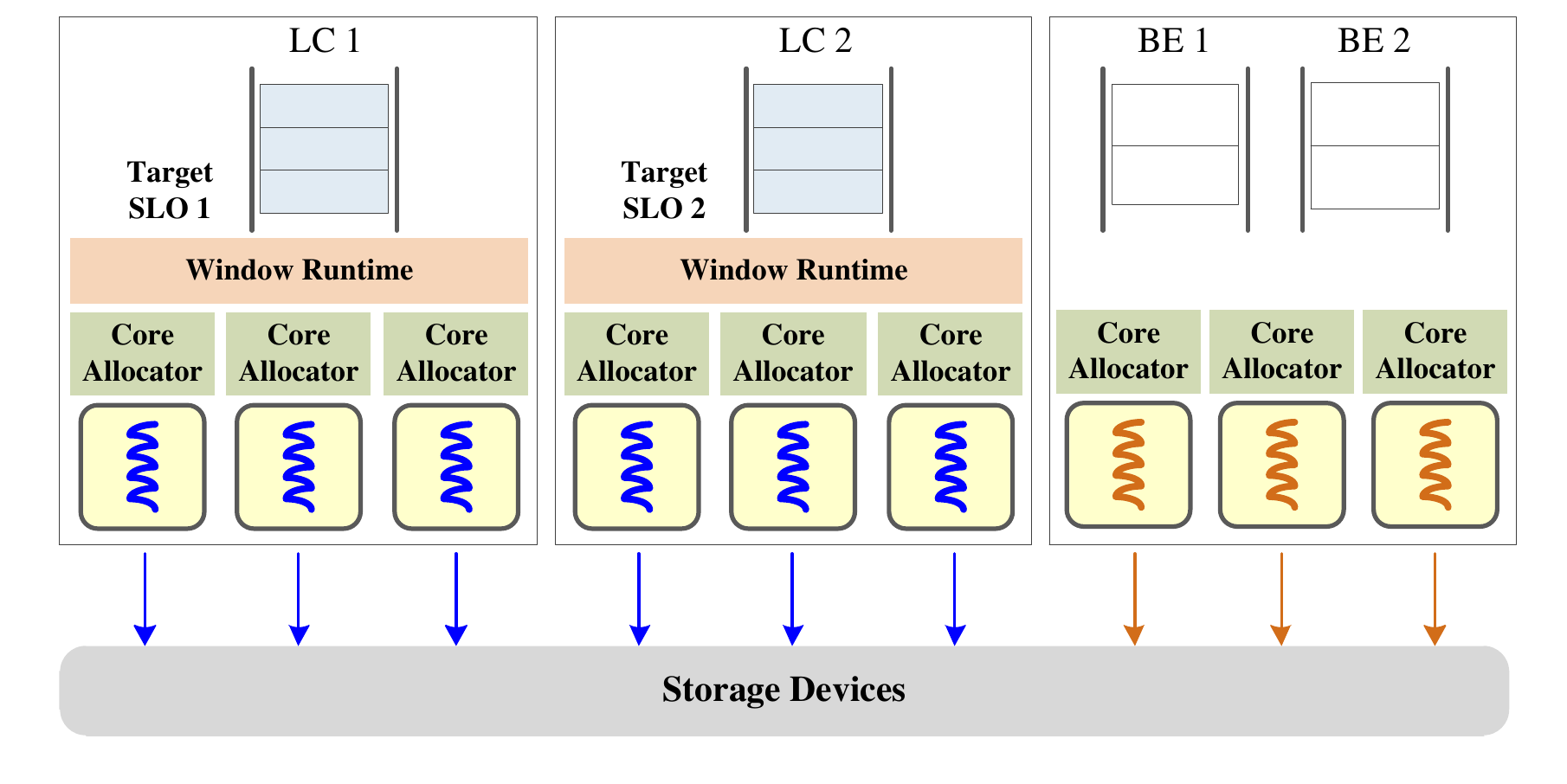}}
\caption{The QWin architecture.}
\label{QWin}
\end{figure}

The goal of QWin is to enforce tail latency SLO of multiple LC tenants while maximize bandwidth of BE tenants that share the storage backend. 
The key idea of QWin is to precisely calculate the number of cores according to target SLO, and dynamically adjust cores.
Figure~\ref{QWin} shows the overall architecture of QWin.
QWin supports multiple levels of target SLO, and each level is corresponded to an LC tenant type which represents a group of LC tenants with a same target SLO. Requests from an LC tenant type are put into a same LC queue. 
Note that an LC tenant hereafter represents for an LC tenant type. 
Request from BE tenants are put into their respective queues.

QWin consists of two components.
\textit{Window Runtime} divides each LC queue into variable-length request windows and collects runtime information (e.g. the number of requests and the waiting time) for each window. 
The number of cores used for each window is accurately calculated according to target SLO and runtime load. 
\textit{Core Allocator} dynamically allocates cores to LC tenants, and the remaining cores are used by BE tenants. 
During processing requests in each window, cores used for LC tenants may be adjusted due to frequently changing interference and loads.

Additionally, each tenant is designated as LC or BE. BE tenants operate at a lower priority: they are only allocated cores that LC tenants do not need. LC tenants will yield extra cores voluntarily when they do not need. If LC tenants need more cores, cores used for BE tenants will be preempted. Before taking a core from BE tenants, the current request handled by the core should be completed. Whereas, cores used by an LC tenant can never be preempted by other tenants.

\subsection{Window Runtime}
\subsubsection{Request-based window}
\label{div-win}
In order to enforce target SLOs and maximize resource utilization, cores must be dynamically allocated since the load is constantly changing with time~\cite{arachne18,shenango19,perfiso18,caladan20,heracles15,parties19,cake12,reflex17}.
The common way to dynamically allocate cores is based on interval~\cite{arachne18,shenango19,caladan20,heracles15,parties19,cake12}.
Although a fixed-time interval is simple, the load in each interval is not definitive because the processing of requests can be across multiple consecutive intervals.
The non-constant load makes it hard to directly determine core requirements. While for a variable-time interval, it is difficult to set a proper interval due to high variable loads. 
Therefore, QWin adopts a flexible variable-length division based on current queuing requests, called request-based window. The size of each window (i.e. the number of requests in each window) can be different, but the runtime load in each window is definitive. 
Once the runtime load is determined, core requirements can be calculated by combining target SLO.

\begin{figure}[htbp]
\setlength{\abovecaptionskip}{0.1cm}
\centerline{\includegraphics[width=0.5\textwidth,trim=60 560 230 150,clip]{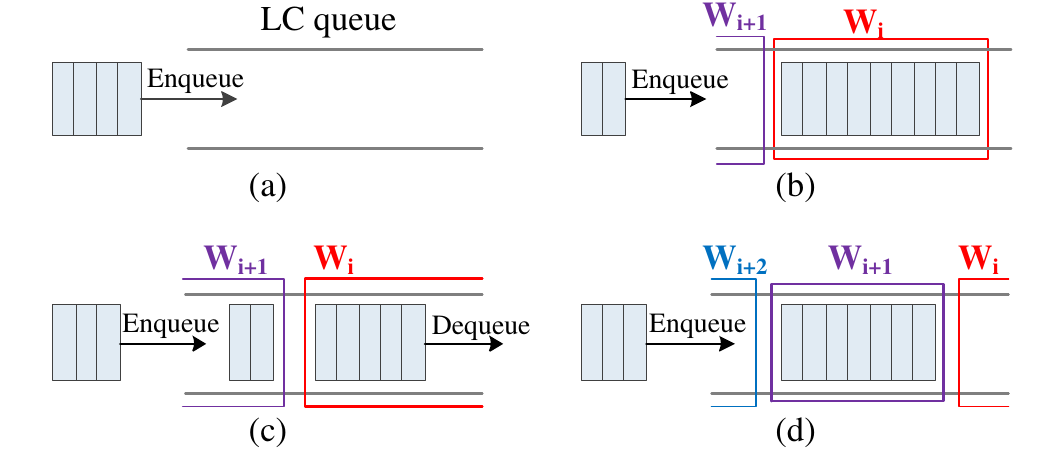}}
\caption{Request-based windows.}
\label{q-win}
\end{figure}

Taking an LC queue as an example, Figure \ref{q-win} shows the method to divide its queue into windows. 
Initially, the queue is empty.
Requests will enter into its queue later (Figure \ref{q-win}(a)).
Then a window $W_i$ is established and all requests in the current queue belong to $W_i$ (Figure \ref{q-win}(b)). If there are $i$ requests in the current queue, the size of $W_i$ is $i$. 
During processing requests in $W_i$, subsequent incoming requests belong to the next window $W_{i+1}$ (Figure \ref{q-win}(c)). 
When all requests in $W_i$ have been handled, $W_i$ ends.
Window $W_{i+1}$ is established when $W_i$ ends, and all requests in the current queue belong to $W_{i+1}$ (Figure \ref{q-win}(d)).
If there are still no incoming requests when a window ends, the queue will be empty. A new window is established until requests enter into the queue again.

In such way, each LC queue is divided into variable-length request-based windows as the changing load. The number of requests in each window reflects the definitive runtime load.

\subsubsection{SLO-to-core calculation model}
\label{est-core}
If a target SLO is guaranteed in each window, the target SLO will be finally guaranteed. Therefore, QWin tries to guarantee target SLOs of LC tenants in each window by dynamically adjusting cores. For each window, QWin introduces the SLO-to-core calculation model to translate a target SLO into the number of cores.

For $W_i$, $QL_i$ is the total number of requests, and $TW_i$ is the queuing time of its first request. $QL_i$ reflects the runtime load, and $TW_i$ reflects the waiting time of $W_i$. 
Let $T_{slo}$ is a target SLO and $Tail_{io}$ is the tail latency of service time (i.e. the I/O latency on storage device of a request), respectively. 
In order to guarantee the target SLO of $W_i$, all $QL_i$ requests must be dequeued within $T_{slo}-Tail_{io}-TW_i$. 
Then the average dequeue rate ($\overline {DR_i}$) of $W_i$ can be calculated by the following formula:
\begin{equation}
\overline {DR_i} = \frac{QL_i}{T_{slo}-Tail_{io}-TW_i}  \label{dr}  
\end{equation}
As long as the real dequeue rate is no less than $\overline {DR_i}$, the tail latency of all requests of $W_i$ cannot exceed target SLO. 
Now let's consider the classic theory Little's Law~\cite{little11} in each window, and let $\overline{T_{io}}$ is the average service time.
Once $W_i$ is established, the number of cores ($N_i$) used to guarantee the target SLO can be calculated as follows:
\begin{equation}
N_i = \overline {DR_i} \times \overline {T_{io}}   \label{n}  
\end{equation}
Both $\overline{T_{io}}$ and $Tail_{io}$ can be obtained in real time while the system is running.
The SLO-to-core calculation model combines target SLO and runtime load to directly calculate the number of cores for LC tenants, while any spare cores can be used by BE tenants.

\subsection{Core Allocator}

\subsubsection{Core policies}
\label{core-policy}
Although SLO-to-core model can be used to precisely calculate the number of cores, only adjusting cores at the begging of each window is not enough for enforcing target SLOs of LC tenants as well as maximizing bandwidth of BE tenants. 
Bursty loads and fluctuated service time that happen during the processing of a window can cause abrupt changes of core usage. All these cases must be promptly detected and the abrupt changes of core usage must be satisfied immediately, otherwise SLO violation happens. That is to say, in order to guarantee target SLO, it is necessary to adjust cores more than once within a processing window.

For example, if a bursty load occurs during the processing of $W_i$, bursty requests belong to $W_{i+1}$ and make it an enormous window. If requests in $W_i$ are not speeded up, the waiting time of requests in $W_{i+1}$ will be longer. Then the tail latency of $W_{i+1}$ may exceed target SLO. Similarly, if the service time of a request in $W_i$ is fluctuated beyond the normal, the queuing time of following requests in $W_i$ will be longer. Even if the number of cores is adjusted at the beginning of $W_{i+1}$, the adjustment is too late for $W_i$ and results in SLO violation. 

Besides, LC tenants with different target SLOs may contend cores with each other. For an LC tenant, it is possible that the number of cores calculated at the beginning of $W_i$ is more than the available cores (including cores that can be preempted from BE tenants). The reason is that other LC tenants temporarily occupies more cores due to the change of load. This leads that the LC tenant cannot get enough cores at the beginning of $W_i$, and it needs to adjust cores again within $W_i$.  

\begin{figure}[htbp]
\setlength{\abovecaptionskip}{0.1cm}
\centerline{\includegraphics[width=0.5\textwidth,trim=60 655 240 95,clip]{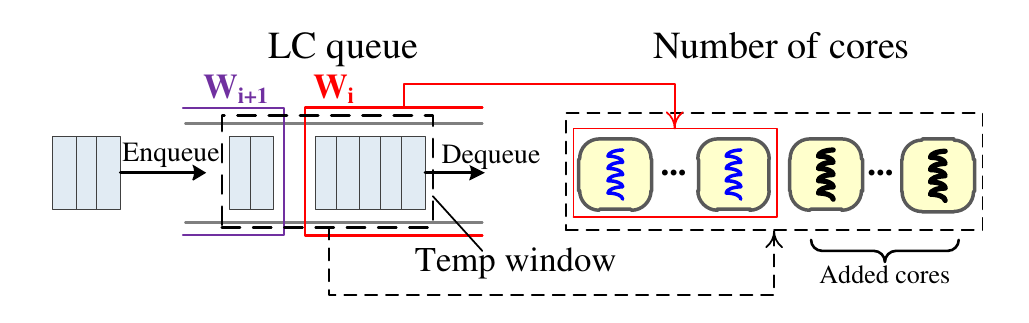}}
\caption{Re-allocate cores for window $W_i$.}
\label{vwin}
\end{figure}

In any case, insufficient cores cause queue to grow. Thus, the change of queue length indicates the changing need of cores.
To deal with the changing need of cores, QWin constantly monitors queue length and adjusts cores within each window.
A temp window is used to reflect the change of queue caused by exceptions that happen within a window.
As shown in Figure \ref{vwin}, at moment $T_i$ during the processing of $W_i$, all requests in current queue belong to a temp window.
Using SLO-to-core model, the corresponding number of cores ($N_t$) for the temp window is calculated. 
If $N_t$ is larger than $N_i$ (the current number of cores), it means that cores are insufficient to enforce target SLO. And more cores ($N_t-N_i$) should be added to promptly process these stacked requests.

In theory, a temp window can be used to check the changing need of cores after processing a request. But, this is not always necessary since these exceptions do not happen at all times, and frequent detection may bring extra overhead. 
Different SLOs reflect different requirements of cores. Tail latency always suffers due to frequently changing interference and load.
It is not enough to respond to all these situations only with one policy.
Therefore, three policies are proposed to adjust cores without compromising SLO within a window: 1) \textit{conservative policy}, which only adjusts cores at the beginning of each window; 2) \textit{aggressive policy}, which continuously adjusts cores after dequeuing a request; 3) \textit{SLO-aware policy}, which adjusts cores after dequeuing \textit{budget} requests. The difference among three policies is the frequency to adjust cores within a window. Obviously, the \textit{budget} used in \textit{conservative policy} and \textit{aggressive policy} is $0$ and $1$, respectively. For \textit{SLO-aware policy}, LC tenants with different SLO should have different \textit{budget}. We use target SLO and the waiting time of $W_i$ to calculate the \textit{budget} for each window:
\begin{equation}
budget = \frac{T_{slo}-Tail_{io}-TW_i}{\overline {T_{io}}}  \label{budget}  
\end{equation}

Tail latency is meaningful based on sufficient requests. QWin polls tail latency of an LC tenant every \texttt{THRESH\_WIN} windows. 
Then, QWin dynamically selects a core policy by monitoring the slack, the difference between target SLO and the current measured tail latency. 
If the slack is more than \texttt{THRESH\_HIGH}, \textit{conservative policy} is set. If the slack is less than \texttt{THRESH\_LOW}, \textit{aggressive policy} is set. Otherwise, \textit{SLO-aware policy} is set. All the tunable parameters are empirically set for our tests; these are described in \cref{eval}.

\subsubsection{Autonomous core allocation} 
Different from core allocation approaches which need an extra core to adjust cores at fixed intervals~\cite{cake12,shenango19,caladan20,heracles15,parties19,arachne18}, no dedicated core is required for core allocation since each core is autonomous in QWin. 
The reason is that each core knows the whole information of tenants it served, including the current window, the current queue, the current core policy, the number of cores and the used cores, etc., making it adjust cores autonomously. Only the cores used by LC tenants can execute algorithm to adjust cores. The cores used by BE tenants depend on the core usage of LC tenants.

Take an LC tenant \textit{L} as example, the core that handles the first request of a window will re-calculate the number of cores, compare it to the current number, and adjust cores when necessary.
If \textit{L} needs more cores, extra cores will be preempted from BE tenants. If \textit{L} occupies too much cores, superfluous cores will be returned to BE tenants.
For \textit{conservative policy}, core adjustment only happens at the beginning of each window.
For \textit{aggressive policy} and \textit{SLO-aware policy}, cores may be adjusted at different frequencies within a window. A core that handles the right request under \textit{aggressive policy} and \textit{SLO-aware policy} will execute core adjustment.
Different from adjusting cores at the beginning of a window, a temp window that can reflect bursty load and fluctuated service time is used to calculate the number of cores within a window when necessary. Note that, in order to avoid unnecessary jitter, core adjustment only monotonously increases within a window. Similarly, increased cores for LC tenants will be preempted from BE tenants. 
To avoid SLO violation, cores used for LC tenants are never preempted in any case.

\subsection{Put Everything Together}
QWin's core allocation runs on each LC core and is integrated with request processing. 
Algorithm \ref{qwin-al} shows the algorithm.
Note that the LC tenant in the algorithm represents a group of LC tenants with a same target SLO. When an LC tenant is allowed to register in the system, QWin allocates a core for it and sets \textit{aggressive policy} as its default core policy. 
After that, the core begins to monitor the LC tenant and its queue, and dynamically adjust cores by itself.

\begin{algorithm}[htbp]
\caption{Core Allocation Algorithm}
\label{qwin-al}
\While{True}{
	t = current\_tenant()\;
	\uIf{t is LC} {
		\uIf{t.win is empty}{
			t.win = new\_win(t.queue)\;
			\uIf{t.wid \% \texttt{THRESH\_WIN} == 0}{
				update\_core\_policy(t)\;
			}
			demand = calculate\_cores(t.win)\;
			adjust\_cores(t, demand)\;
		}
		handle a request from t in FIFO order\;
		t.wcnt++\;
		\uIf{t.win is not empty \&\& t.budget != 0 \&\& t.wcnt \% t.budget == 0}{
			tw = temp\_win(t.queue)\;
			tmp = calculate\_cores(tw)\;
			\uIf{tmp > t.num}{
				adjust\_cores(t, tmp)\;
			}
		}
		\uIf{t.win is empty \&\& t.queue is empty \&\& t.num > 1}{
			yield this core to BE tenants\;
			t.num -= 1\;
		}
	}
}
\end{algorithm}

For any core, the tenant it served is checked. If the tenant is LC, the algorithm is executed as follows.
First, if the current window is empty, a new window will be established. \texttt{calculate\_cores()} is call to calculate the number of cores by Formula~\ref{n}, and \texttt{adjust\_cores()} is called to adjust cores (lines 4-9). Meanwhile, the core periodically checks if core policy should be changed by a configurable threshold, and \texttt{update\_core\_policy()} is called to adjust core policy based on the method mentioned in \cref{core-policy} (lines 6-7)
Second, the thread running on this core will handle a request from the LC queue, and \texttt{t.wcnt}, the number of completed requests, is updated (lines 10-11).
Third, the core checks if the number of cores need to be adjusted according to current core policy (except for \textit{conservative policy}, which only adjusts core at the beginning of each window) (lines 12-16). Core policies are distinguished by different \texttt{budget} as mentioned in \cref{core-policy}. If the current window is not empty and \texttt{t.wcnt} is divisible by \texttt{budget}, a temp window is established to calculate the current need of cores, and if more cores are needs, \texttt{adjust\_cores()} is called to add cores. 
Finally, if the current window is empty and the queue is also empty, the core will be yielded to BE tenants unless it is the last core used for this LC tenant (lines 17-19).

\begin{algorithm}
\caption{QWin's \textit{adjust\_cores(t, target)} function}
\label{adjust}
\tcp{@t is an LC tenant}
\tcp{@target is the target number of cores}
\uIf{t.num > target}{
	yield \textit{t.num - target} cores to BE tenants\;
	t.num = target\;
}
\uElseIf{t.num < target}{
	delta = target - t.num\;
	available = get\_BE\_cores()\;
	n = delta < available ? delta : available\;
	preempt $n$ cores from BE tenants to t\;
	t.num += n\;
}
\end{algorithm}

The core adjustment function \texttt{adjust\_cores()} is shown in Algorithm \ref{adjust}. For an LC tenant \textit{t}, if the current number of cores, \texttt{t.num}, is larger than the target number of cores, \texttt{target}, superfluous cores (\texttt{t.num}-\texttt{target}) are yielded to BE tenants (lines 1-3). Otherwise, more cores will be preempted from BE tenants and assigned to this LC tenant (lines 4-9). If it hanppens that LC tenant cannot be assigned enough cores (\texttt{delta} $>$ \texttt{available}), subsequent core allocation will adjust cores again (as mentioned in \cref{core-policy}).

\section{Implementation}
\label{impl}
We implement QWin in the storage backend (OSD) of the widely-used open-source release of Ceph, but a completely new core allocation strategy is implemented. Moreover, QWin can be integrated into any storage backend which adopts the classical SEDA architecture. Meanwhile, QWin does not need any offline profiling, or a priori knowledge about each tenant's characteristics except for their target SLOs, making it applicable in wider scenarios.

We modify Ceph's OSD in three important ways. 
First, we add a registration module in OSD to receiving target SLOs (e.g., 99.9th percentile latency) of LC tenants. 
Second, we create a separate FIFO queue for each tenant to reduce interference from contention of a single queue, since that previous works~\cite{tales14,zygos17,adam12} has analyzed that FIFO queuing is the lowest tail latency.  QWin's \textit{Window Runtime} monitors each queue and quantifies the runtime load of each LC tenant by the flexible variable-length windows.
Third, an autonomous core allocation strategy is integrated into the processing of request scheduling. The strategy allows us to completely eliminate the waste of a dedicated core to adjust cores imposed by other systems~\cite{cake12,shenango19,caladan20,perfiso18}.
Besides, in order to guarantee target SLOs of tenants accessing the storage backend, we modify OSD to enforce access control list (ACL) policies at the granularity of tenants. It checks if a tenant has the right to access the system during the registration. If the tenant is not permitted, it cannot open a connection to the storage backend.

\section{Evaluation}
\label{eval}
In evaluating QWin, we aim to answer the following questions:

1. How can QWin enforcing target SLOs for LC tenants meanwhile maximizing bandwidth of BE tenants compare to previous systems? (\cref{Q-Perf})

2. Can QWin satisfy diverse target SLOs for multiple LC tenants? (\cref{p2})

3. How do three core policies provided by QWin enable it to perform well? (\cref{p3})

\begin{table}[]
\caption{Workloads used in our evaluation.}
\label{workload}
\begin{tabular}{|c|l|l|c|}
\hline
Label & \multicolumn{2}{c|}{Characterization}                                                                                                                       & LC/BE \\ \hline
A     & \multirow{4}{*}{\begin{tabular}[c]{@{}l@{}}Fio configure:\\ bs=4KB;\\ iodepth=16;\\ numjobs=8;\end{tabular}}  & randread;                                   & LC    \\ \cline{1-1} \cline{3-4} 
B     &                                                                                                               & \multicolumn{1}{c|}{randrw;readratio=95\%;} & LC    \\ \cline{1-1} \cline{3-4} 
C     &                                                                                                               & \multicolumn{1}{c|}{randrw;readratio=90\%;} & LC    \\ \cline{1-1} \cline{3-4} 
D     &                                                                                                               & randrw;readratio=85\%;                      & LC    \\ \hline
E     & \multirow{4}{*}{\begin{tabular}[c]{@{}l@{}}Fio configure:\\ bs=64KB;\\ iodepth=16;\\ numjobs=2;\end{tabular}} & read;                                   & BE    \\ \cline{1-1} \cline{3-4} 
F     &                                                                                                               & rw;readratio=99\%;                      & BE    \\ \cline{1-1} \cline{3-4} 
G     &                                                                                                               & rw;readratio=95\%;                      & BE    \\ \cline{1-1} \cline{3-4} 
H     &                                                                                                               & rw;readratio=90\%;                      & BE    \\ \hline
J     & \multicolumn{2}{l|}{OLTP from Filebench;}                                                                                                                   & LC    \\ \hline
K     & \multicolumn{2}{l|}{Webserver from Filebench;}                                                                                                              & LC    \\ \hline
P     & \multicolumn{2}{l|}{\begin{tabular}[c]{@{}l@{}}Fio configure:bs=4KB;iodepth=32;\\ numjobs=8;randrw;readratio=90\%;\end{tabular}}                            & LC    \\ \hline
\end{tabular}
\end{table}

\captionsetup[subfigure]{labelformat=empty}
\begin{figure*}[] 
  \centering
  \setlength{\abovecaptionskip}{5pt}
  \subfigbottomskip=0pt
  \subfigcapskip=-5pt
   \subfigure{
    \includegraphics[width=0.8\textwidth,trim=2 130 8 80,clip]{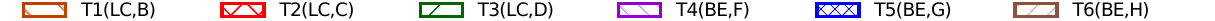}}
   
   \subfigure{
    \includegraphics[width=0.24\textwidth,trim=2 58 190 96,clip]{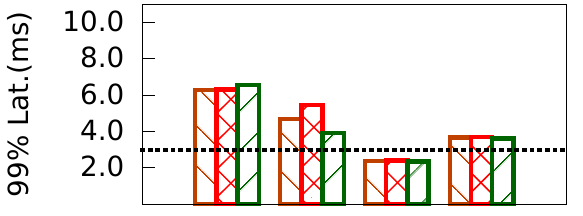}}
  \subfigure{
    \includegraphics[width=0.24\textwidth,trim=2 58 190 96,clip]{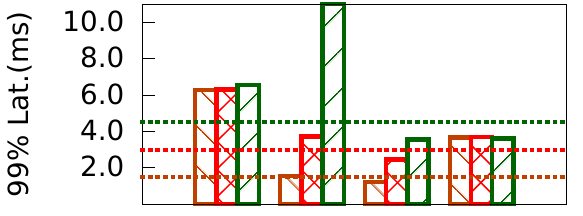}}
    \subfigure{
    \includegraphics[width=0.24\textwidth,trim=2 58 190 96,clip]{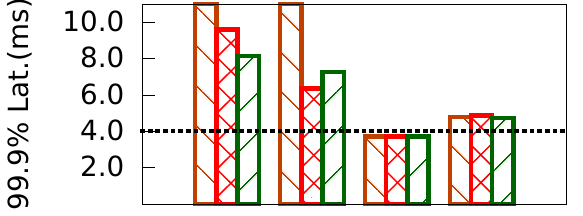}}
     \subfigure{
    \includegraphics[width=0.24\textwidth,trim=2 58 190 96,clip]{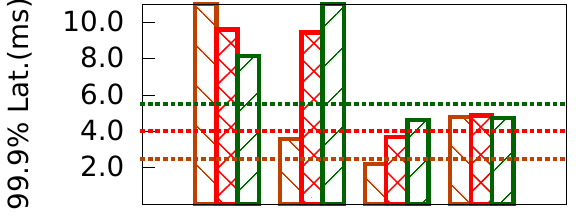}}
    
    \setcounter{subfigure}{0}
    \subfigure[SLO(99\% Lat.): 3ms]{
    \includegraphics[width=0.24\textwidth,trim=2 45 190 95,clip]{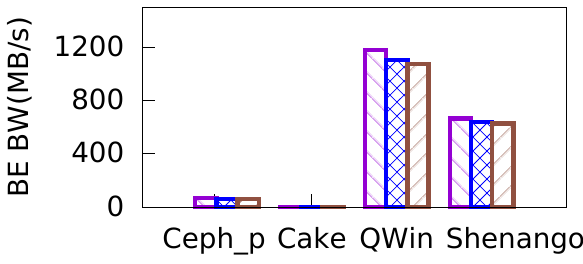}}
  \subfigure[SLO(99\% Lat.): 1.5/3/4.5ms]{
    \includegraphics[width=0.24\textwidth,trim=2 45 190 95,clip]{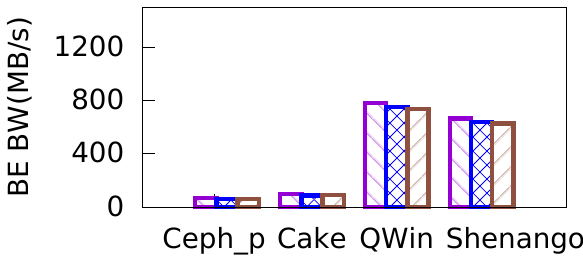}}
    \subfigure[SLO(99.9\% Lat.): 4ms]{
     \includegraphics[width=0.24\textwidth,trim=2 45 190 95,clip]{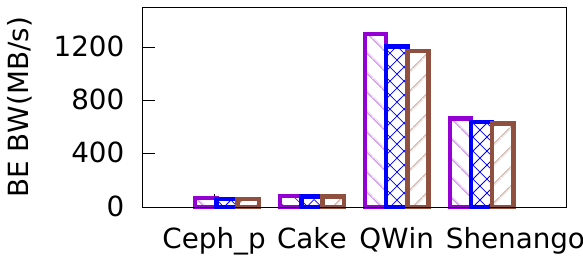}}
 \subfigure[SLO(99.9\% Lat.): 2.5/4/5.5ms]{
    \includegraphics[width=0.24\textwidth,trim=2 45 190 95,clip]{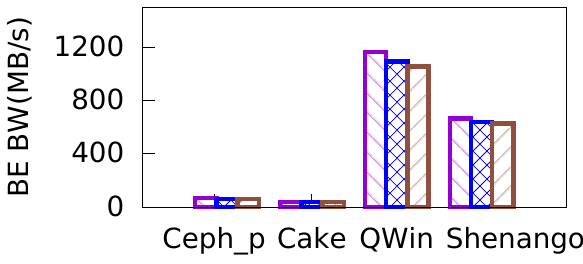}}
  \caption{QWin can consolidate six tenants (T1(LC, B), T2(LC, C), T3(LC, D), T4(BE, F), T5(BE, G), T6(BE, H)) with satisfying target SLOs of LC tenants (top), while maintaining excellent bandwidth for BE tenants (bottom). (a) and (c) show a same target SLO (99th/99.9th) for three LC tenants (the black dashed line is target SLO); while (b) and (d) show three respective target SLOs (99th/99.9th) for three LC tenants (each of the colored horizontal dashed lines correspond to the SLO of the similarly colored tenant).}
  \label{fio} 
\end{figure*}

\textbf{Experimental Setup.}
Our testbed consists of eleven machines, and each has dual 12 physical core Intel Xeon E5-2650v4 CPU, 64GB RAM, two 10GbE NIC, and two 1TB Intel P4510 SSDs, running CentOS 7.3.  
The Ceph cluster (Ceph 10.2.11 with the BlueStore backend) consists of one monitor and eight OSDs. The monitor is running on a single machine.
Two OSDs are co-located on a same machine, and each OSD is configured with a 10GbE NIC, an SSD and a dual 12-core physical CPU. 
We use six machines as clients, three for LC tenants and three for BE tenants. A 30GB RBD~\cite{rbd19} image is created in the Ceph cluster for each tenant. Both Fio~\cite{fio} and Filebench~\cite{filebench} are used to simulate the workload of LC and BE tenants. The characterizations of all workloads used in the following experiments are shown in Table~\ref{workload}. 
In order to clarify the configuration of tenants in each experiment, we use T1(LC, A) for short to describe a LC tenant issuing workload A, and T2(BE, E) to describe a BE tenant issuing workload E.

\textbf{Compared Systems.}
We compare QWin to Ceph-priority (Ceph\_p for short) which can distinguish LC and BE tenants by setting different priorities, Cake which is an SLO-aware approach but proportional sharing cores, and Shenango which is an SLO-unaware approach but using trial-and-error core adjustment. 
Since Ceph\_p can only support two coarse-grained priorities (high and low), we configure high priority for all LC tenants and low priority for all BE tenants. And then LC requests are always processed first.
We implement Cake and Shenango in Ceph's OSD.
Cake~\cite{cake12} works by dynamically adjusting proportional shares and reservations of threads (assuming that each thread uses a dedicated core) to meet target SLOs. Since Cake only supports a single LC tenant and a BE tenant, we extend Cake's SLO compliance-based scheduling algorithm to support multiple LC and BE tenants. 
Shenango~\cite{shenango19} dynamically adjusts cores based on queued requests of LC tenants. If a request is present in the queue for two consecutive intervals, an extra core is added. When the queue is empty, all allocated cores will be reclaimed at once. 

\textbf{Parameters.} QWin has three tunable parameters. In our evaluation, we empirically set \texttt{THRESH\_WIN}, \texttt{THRESH\_LOW}, \texttt{THRESH\_HIGH} to $2000$ windows, $300 \mu s$ and $1000 \mu s$, respectively. The choice of an SLO is driven by LC tenant's requirement and load, with the intuitive understanding that a more stringent SLO requires more resources. How to determine a good SLO is out of the scope of this paper. The SLO used in our evaluation is tentative. In each experiment, we vary the adjustment interval for Cake ($1s$, $5s$ and $10s$) and Shenango (intervals are shown in Figure~\ref{shenango}), and both choose the interval that the tail latencies of LC tenants are the lowest.

\captionsetup[subfigure]{labelformat=empty}
\begin{figure*}[h] 
  \centering
  \setlength{\abovecaptionskip}{5pt}
  \subfigbottomskip=0pt
  \subfigcapskip=-5pt
   \subfigure{
    \includegraphics[width=0.8\textwidth,trim=2 130 8 80,clip]{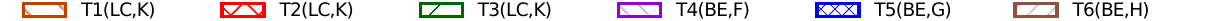}}
   
   \subfigure{
    \includegraphics[width=0.24\textwidth,trim=2 58 190 96,clip]{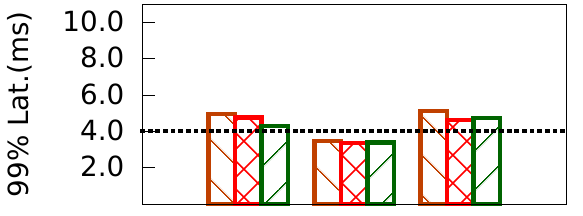}}
  \subfigure{
    \includegraphics[width=0.24\textwidth,trim=2 58 190 96,clip]{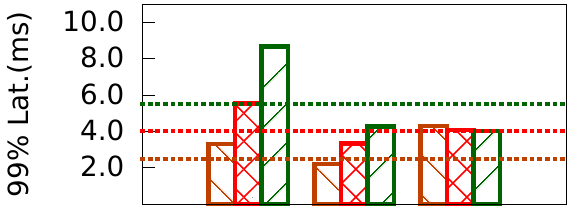}}
    \subfigure{
    \includegraphics[width=0.24\textwidth,trim=2 58 190 96,clip]{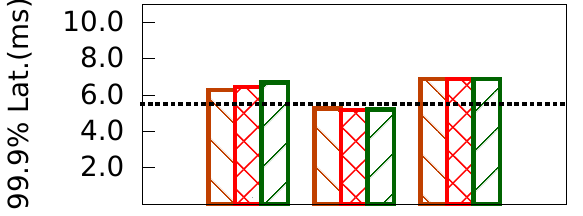}}
     \subfigure{
    \includegraphics[width=0.24\textwidth,trim=2 58 190 96,clip]{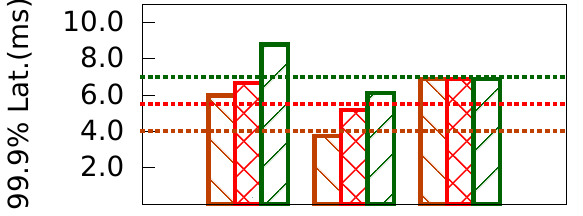}}
    
    \setcounter{subfigure}{0}
    \subfigure[SLO(99\% Lat.): 4ms]{
    \includegraphics[width=0.24\textwidth,trim=2 45 190 95,clip]{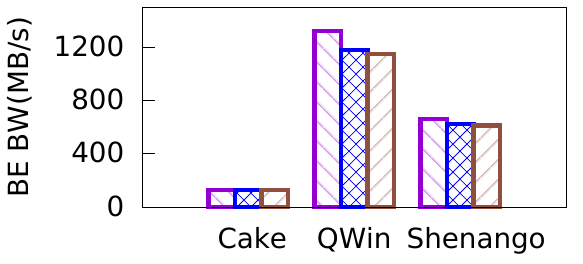}}
  \subfigure[SLO(99\% Lat.): 2.5/4/5.5ms]{
    \includegraphics[width=0.24\textwidth,trim=2 45 190 95,clip]{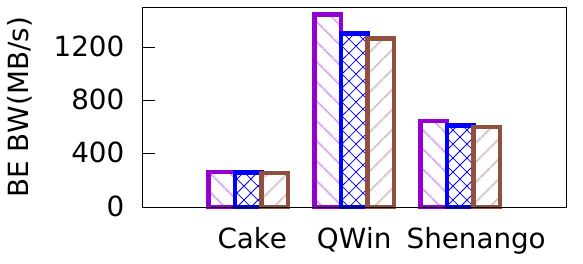}}
    \subfigure[SLO(99.9\% Lat.): 5.5ms]{
     \includegraphics[width=0.24\textwidth,trim=2 45 190 95,clip]{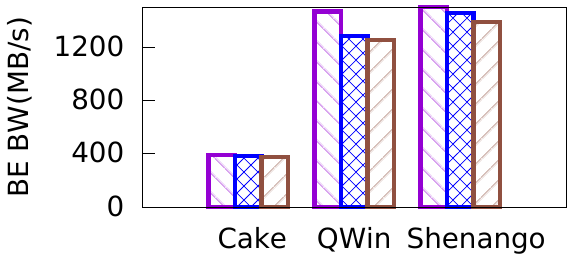}}
 \subfigure[SLO(99.9\% Lat.): 4/5.5/7ms]{
    \includegraphics[width=0.24\textwidth,trim=2 45 190 95,clip]{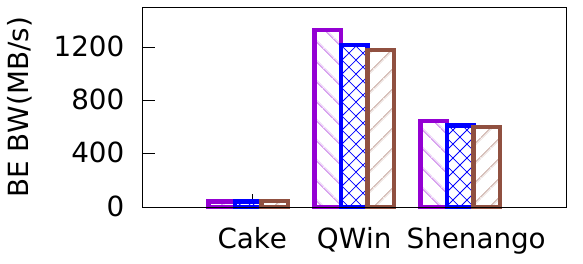}}
  \caption{QWin can consolidate three LC tenants (each issues Webserver, T1(LC, K), T2(LC, K), T3(LC, K)) and three BE tenants (T4(BE, F), T5(BE, G), T6(BE, H)) with satisfying target SLOs of LC tenants (top), while maintaining excellent bandwidth for BE tenants (bottom).}
  \label{webserver} 
\end{figure*}

\captionsetup[subfigure]{labelformat=empty}
\begin{figure*}[h] 
  \centering
  \setlength{\abovecaptionskip}{5pt}
  \subfigbottomskip=0pt
  \subfigcapskip=-5pt
   \subfigure{
    \includegraphics[width=0.8\textwidth,trim=2 130 8 80,clip]{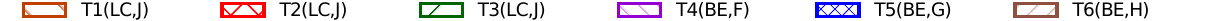}}
   
   \subfigure{
    \includegraphics[width=0.24\textwidth,trim=2 58 190 96,clip]{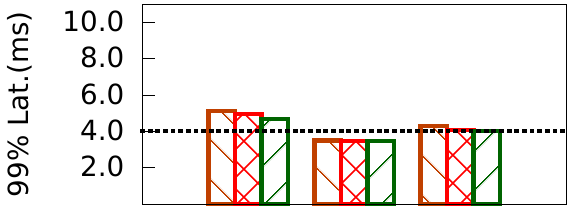}}
  \subfigure{
    \includegraphics[width=0.24\textwidth,trim=2 58 190 96,clip]{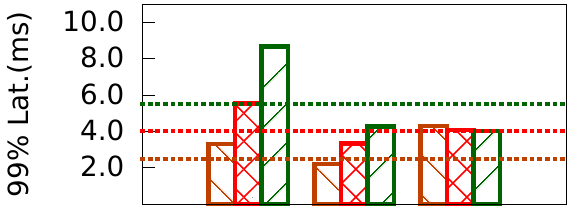}}
    \subfigure{
    \includegraphics[width=0.24\textwidth,trim=2 58 190 96,clip]{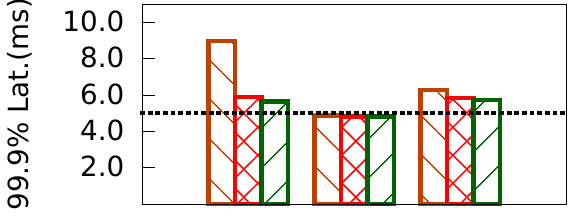}}
     \subfigure{
    \includegraphics[width=0.24\textwidth,trim=2 58 190 96,clip]{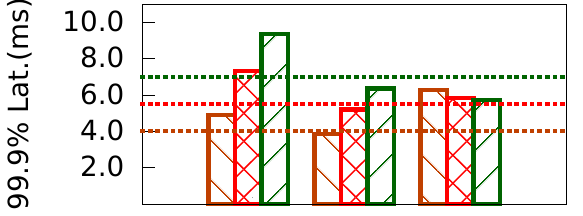}}
    
    \setcounter{subfigure}{0}
    \subfigure[SLO(99\% Lat.): 4ms]{
    \includegraphics[width=0.24\textwidth,trim=2 45 190 95,clip]{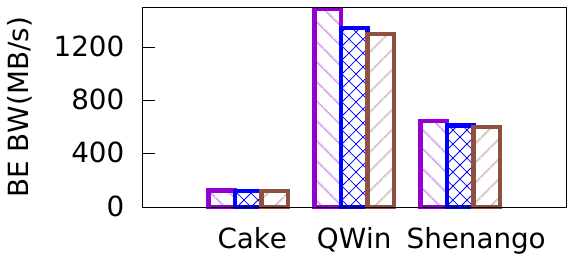}}
  \subfigure[SLO(99\% Lat.): 2.5/4/5.5ms]{
    \includegraphics[width=0.24\textwidth,trim=2 45 190 95,clip]{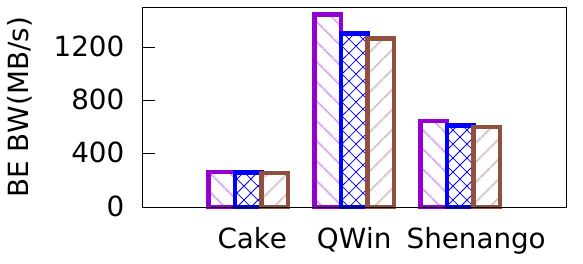}}
    \subfigure[SLO(99.9\% Lat.): 5ms]{
     \includegraphics[width=0.24\textwidth,trim=2 45 190 95,clip]{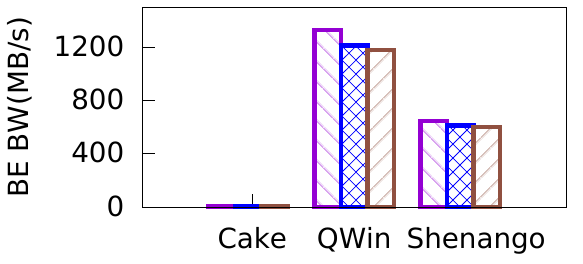}}
 \subfigure[SLO(99.9\% Lat.): 4/5.5/7ms]{
    \includegraphics[width=0.24\textwidth,trim=2 45 190 95,clip]{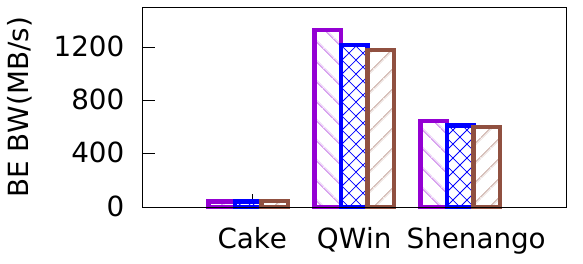}}
  \caption{QWin can consolidate three LC tenants (each issues OLTP, T1(LC, J), T2(LC, J), T3(LC, J)) and three BE tenants (T4(BE, F), T5(BE, G), T6(BE, H)) with satisfying target SLOs of LC tenants (top), while maintaining excellent bandwidth for BE tenants (bottom).}
  \label{oltp} 
\end{figure*}

\subsection{Enforcing target SLO and Increasing Bandwidth}
\label{Q-Perf}
We now compare QWin with Ceph\_p, Cake and Shenango by evaluating the ability to enforce target SLO for LC tenants while maximize bandwidth for BE tenants. 
Three groups of consolidations were designed. Each group consolidates three LC tenants along with three BE tenants. The LC tenants in three groups are different: 1) three LC tenants (T1, T2 and T3) run three Fio workloads B, C and D in Table~\ref{workload}, respectively; 2) three LC tenants (T1, T2 and T3) all run Webserver from Filebench (workload K in Table~\ref{workload}); 3) three LC tenants (T1, T2 and T3) all run OLTP from Filebench (workload J in Table~\ref{workload}).
Three BE tenants (T4, T5 and T6) in each group run three Fio workloads F, G and H in Table~\ref{workload}, respectively. 
Both 99th tail latency SLO and 99.9th tail latency SLO are evaluated in this section for LC tenants. Meanwhile, in each group, three LC tenants are set with a same SLO, or each LC tenant is set with a respective SLO. 

Results for three groups are shown in Figure~\ref{fio}, Figure~\ref{webserver} and Figure~\ref{oltp}, respectively.
For both Fio simulated LC tenants and typical LC tenants (OLTP and Webserver), QWin is the only one that meets target SLOs (99th/99.9th) for LC tenants and yields excellent BE tenants bandwidth. The bandwidth of BE tenants is about 1.2x$\sim$31x higher than other compared systems without compromising target SLO of LC tenants. 
Cake can hardly satisfy any SLO for LC tenants, and the bandwidth of BE tenants is much lower than QWin. 
Shenango does not consider target SLO to adjust cores. Even if LC tenants with different SLO, the tail latency (99th/99.9th) of each LC tenant is nearly the same, and SLO violations happen. 
Besides, both Cake and Shenango need a dedicated core to adjust cores, which is also a waste of cores. 
Ceph\_p currently only support two priorities: high and low, so it cannot further distinguish LC tenants with different SLO by different priorities. 
Although LC requests have high priority over BE requests, target SLO still cannot be guaranteed. This is because all cores are shared among LC and BE tenants, and the interference cannot be avoided. 

For Cake, either 99th percentile latency or 99.9th percentile latency are not satisfied to corresponding target SLO. During the experiments, we find that Cake only reserves a few cores for each LC tenant, and other cores are shared in proportion among six tenants. 
For LC tenants, although the proportional share is more than 90\%, tail latency (99th/99.9th) is still significantly impacted due to the interference from other tenants. 
The erraticism of tail latency in each interval (as shown in Figure~\ref{cake}) is aggravated when read requests and write requests are accessing the underlying storage devices simultaneously. The erratic historical tail latency leads to a wrong adjustment, making SLO violation happens. 
Moreover, the reactive feedback-control based approach adopted by Cake cannot respond to bursty load immediately, and this also results in SLO violation.
Besides, the overhead of calculating 99.9th percentile latency in each interval is non-trivial. 

For each LC tenant, Shenango only adds a core when congestion is detected, and aggressively acquires cores merely based on runtime load, not target SLO. 
We monitor the core allocation of Shenango, and find that even if the target SLO is looser, it always occupies more cores than its real need, which results in poor bandwidth of BE tenants. When the load suddenly increases or service time is more fluctuated, incremental adjustment of cores makes that LC tenants need more time to acquire enough cores, and this can have a serious impact on tail latency, especially the 99.9th percentile latency. For all experiments, 99th and 99.9th percentile latencies of three LC tenants in same test are nearly the same. The reason is that Shenango does not distinguish LC tenants that have different target SLO. 

For LC tenants with any target SLO (99th/99.9th, same/different), QWin can precisely calculate cores combining with the flexible request-based window, and promptly adjust cores by proper core policy when bursty load and fluctuated service time are detected. From our statistics in the first group of experiment, up to 40,000 variable-length windows are established per minutes to quantify runtime load, and up to 85,000 core allocations are executed per minutes to adjust cores for the changing need of cores.
Meanwhile, QWin adaptively changes core policies for LC tenants to adjust cores to respond to the changing needs (further experiments are shown in \cref{p3}).
QWin's fast and precise reactions enable all three LC tenants to maintain target SLO throughout constantly shifting load and fluctuated service time. 
All the above experiments show that QWin obviously superior than existing approaches in both enforcing target SLO of LC tenants and enhancing bandwidth of BE tenants.

\subsection{Diverse target SLOs}
\label{p2}
To understand if QWin can maintain its benefits under diverse target SLOs, two scenarios are evaluated: 1) read-only (100\% read), where three LC tenants (T1, T2 and T3) all run Fio workload A in Table~\ref{workload} and three BE tenants (T4, T5 and T6) all run Fio workload E in Table~\ref{workload}; 2) read-heavy (more than 85\% read), where three LC tenants (T1, T2 and T3) run three Fio workload B, C and D in Table~\ref{workload} and three BE tenants (T4, T5 and T6) run three Fio workload F, G and H in Table~\ref{workload}.
Note that three different 99.9th percentile tail latency (strict/general/loose) are set as target SLO. There are three tests in each scenario, and in each test, three LC tenants are set with a same target SLO. 
We compare QWin with Cake which adopts a different strategy to adjust cores with considering target SLO.

\captionsetup[subfigure]{labelformat=empty}
\begin{figure}[h]
  \centering
  \setlength{\abovecaptionskip}{5pt}
  \subfigbottomskip=0pt
  \subfigcapskip=-7pt
   \subfigure{
    \includegraphics[width=0.45\textwidth,trim=2 127 100 70,clip]{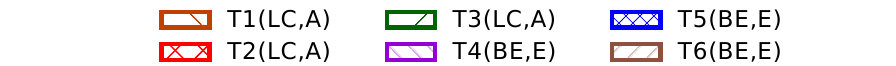}}
    \subfigure{
    \includegraphics[width=0.45\textwidth,trim=4 130 80 75,clip]{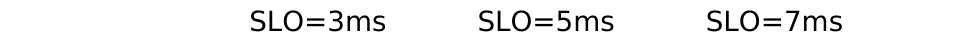}}
   \subfigure{
    \includegraphics[width=0.45\textwidth,trim=4 60 80 96,clip]{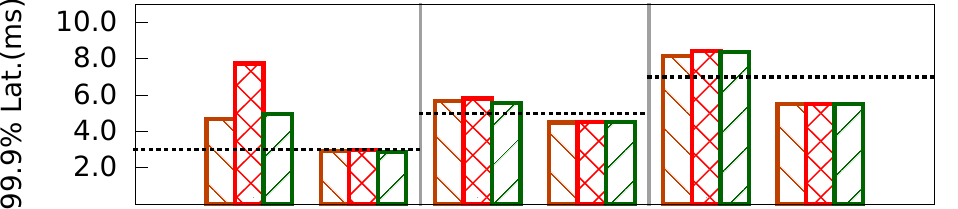}}
     \setcounter{subfigure}{0}
    \subfigure[Scenario 1: read-only]{
    \includegraphics[width=0.45\textwidth,trim=4 40 80 96,clip]{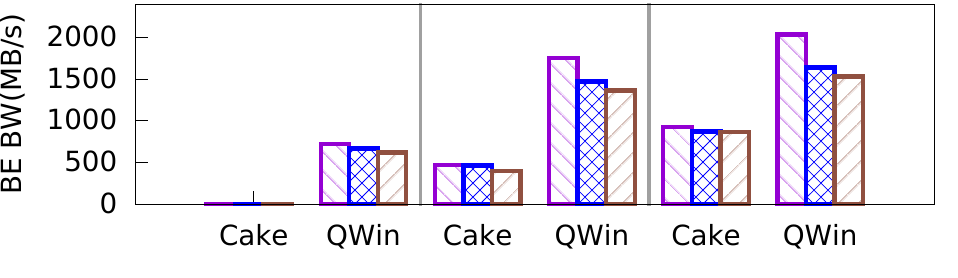}}
    \subfigure{
    \includegraphics[width=0.45\textwidth,trim=2 127 100 70,clip]{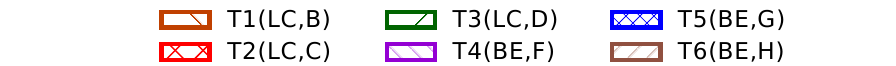}}
    \subfigure{
    \includegraphics[width=0.45\textwidth,trim=4 130 80 75,clip]{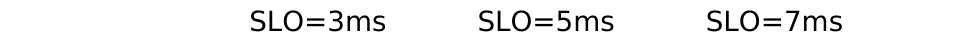}}
   \subfigure{
    \includegraphics[width=0.45\textwidth,trim=4 60 80 96,clip]{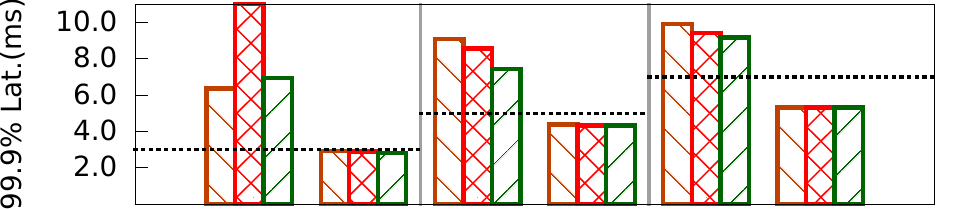}}
     \setcounter{subfigure}{1}
    \subfigure[Scenario 2: read-heavy]{
    \includegraphics[width=0.45\textwidth,trim=4 40 80 96,clip]{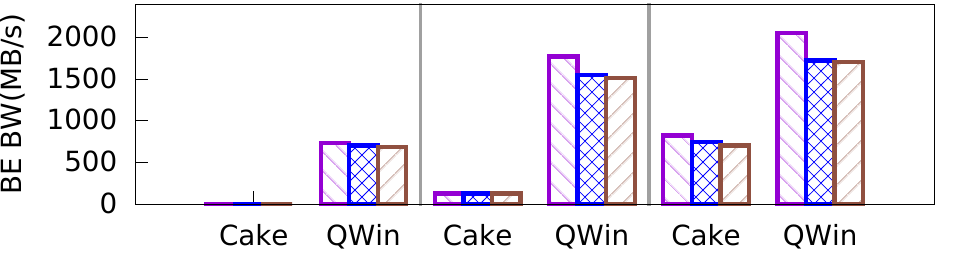}}
  \caption{When consolidate multiple LC and BE tenants, QWin can adapt to diverse target SLOs with satisfying SLOs for LC tenants (top in (a) and (b)), while maintaining excellent bandwidth for BE tenants (bottom in (a) and (b)).}
  \label{dfio} 
\end{figure}

As shown in Figure~\ref{dfio}, QWin is able to guarantee different target SLOs for LC tenants in two scenarios, and as expected, as target SLOs become looser, bandwidth of BE tenants becomes higher. A looser target SLO uses less cores, and then the remaining cores can be fully used by BE tenants. 
At any time, including facing bursty loads and fluctuated service time, QWin can precisely calculate the number of cores, and allocate these cores at a time to meet the changing need, instead of increasing them one by one. 
Differently, Cake only compares the historical tail latency and target SLO to adjust cores. This results in inaccurate core allocation, and it is extremely hard to quickly respond to incoming bursty load or fluctuated service time.
Besides, the reactive feedback control to adjust cores adopted by Cake is always too late to handle the fluctuated service time or bursty load. 
When a strict target SLO (i.e. $3ms$) is set for LC tenants in both two scenarios, even if Cake allocates almost all the cores to LC tenants, target SLO is still not satisfied and the bandwidth of BE tenants is nearly zero. 
Meanwhile, Cake can even hardly satisfy a looser target SLO for LC tenants. 
That is because a tiny sharing of cores can also significantly impact the tail latency.

\subsection{Effects of core policies}
\label{p3}
To evaluate the benefits of combining three core policies in QWin, two LC tenants and a BE tenant are consolidated in storage backend. Two LC tenants (T1 and T2) run two Fio workload C and P in Table~\ref{workload}, respectively, and the BE tenant (T3) runs Fio workload H in Table~\ref{workload}.
We compare target SLOs (99.9th percentile tail latency) of LC tenants, the bandwidth of BE tenant and the allocation of cores under three policies mentioned in \cref{core-policy}: 1) \textit{conservative policy}; 2) \textit{aggressive policy}; 3) \textit{SLO-aware policy}; and QWin which dynamically selects one of three core policies based on the actual situation. The target SLO (99.9th percentile latency) for T1(LC, C) and T2(LC, P) is set to $3ms$ and $5ms$, respectively.

\captionsetup[subfigure]{labelformat=empty}
\begin{figure}[h]
  \centering
  \setlength{\abovecaptionskip}{5pt}
  \subfigbottomskip=0pt
  \subfigcapskip=-7pt
  \subfigure{
    \includegraphics[width=0.45\textwidth,trim=4 125 80 68,clip]{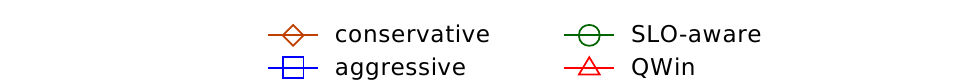}}
   \subfigure{
    \includegraphics[width=0.45\textwidth,trim=4 63 80 93,clip]{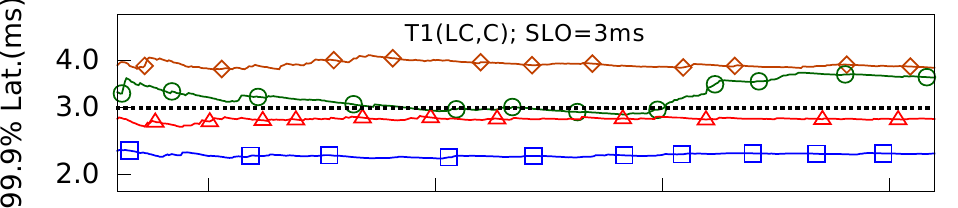}}
    \subfigure{
    \includegraphics[width=0.45\textwidth,trim=4 63 80 93,clip]{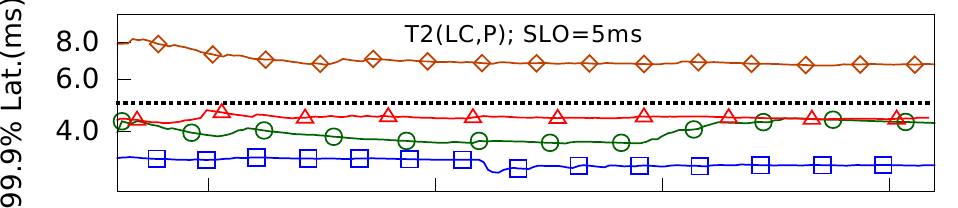}}
    \subfigure{
    \includegraphics[width=0.45\textwidth,trim=4 63 80 97,clip]{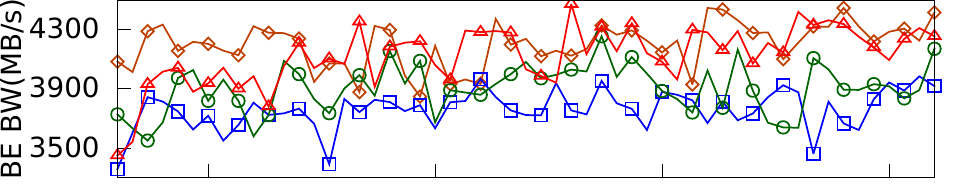}}
   \subfigure{
    \includegraphics[width=0.45\textwidth,trim=4 63 80 97,clip]{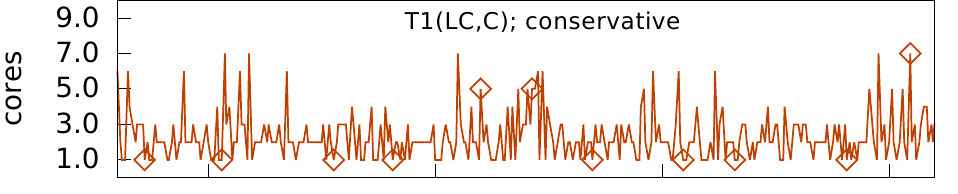}}
    \subfigure{
    \includegraphics[width=0.45\textwidth,trim=4 63 80 97,clip]{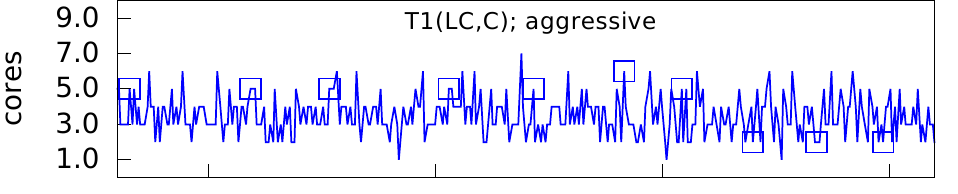}}
    \subfigure{
    \includegraphics[width=0.45\textwidth,trim=4 63 80 97,clip]{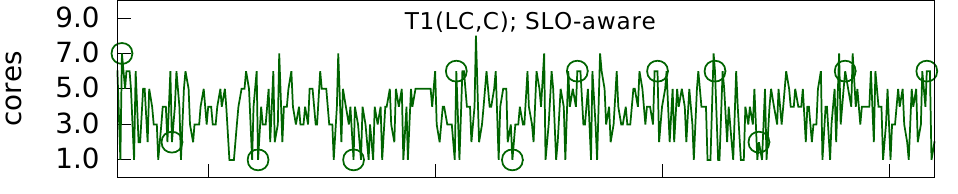}}
    \subfigure{
    \includegraphics[width=0.45\textwidth,trim=4 40 80 97,clip]{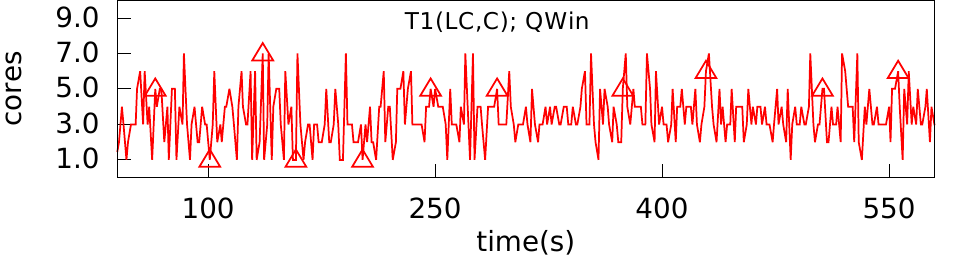}}
  \caption{Benefits of different core policies in QWin.}
  \label{3p} 
\end{figure}

Results of this experiment is shown in Figure~\ref{3p}.
For \textit{conservative policy}, target SLO for two LC tenants cannot be guaranteed. 
For \textit{aggressive policy} and QWin, target SLO for two LC tenants are both satisfied, but under \textit{aggressive policy}, the 99.9th percentile latencies are much lower than target SLOs.
For \textit{SLO-aware policy}, target SLO for T1(LC, C) is not satisfied, while target SLO for T2(LC, P) is satisfied.
The bandwidth of BE tenant, T3(BE, H), is constantly changing during the experiment due to the changing of its available cores. 
We calculate the average bandwidth for four strategies: 4200MB/s under \textit{conservative policy}, 3759MB/s under \textit{aggressive policy}, 3906MB/s under \textit{SLO-aware policy}, and 4116MB/s for QWin.
The experimental results show that for QWin, not only target SLO of LC tenants are satisfied (with the closest 99.9 percentile tail latency to target SLO), but also the bandwidth of BE tenant is highest.

We also monitor the core allocation during this experiment. Only the results of core allocation for T1(LC, C) are shown in Figure~\ref{3p} (lower four graphs). Results for T2(LC, P) are similar but not shown for brevity.
Compare to other policies, more cores are allocated for \textit{aggressive policy}. That is because \textit{aggressive policy} continuously checks and adjusts cores after processing a request. Such policy is too aggressive, and more cores are allocated to LC tenants, seriously decreasing the bandwidth of BE tenants. \textit{conservative policy} only adjusts cores once per window, which can not quickly respond to bursty load and fluctuated service time. The frequency to adjust cores under \textit{SLO-aware policy} is between \textit{aggressive policy} and \textit{conservative policy}. Such policy is related to target SLO, and may result in SLO violation (the top graph) due to late adjustment.
QWin flexibly combines three core policies, and dynamically select a core policy to adjust cores. During the experiment, both T1(LC, C) and T2(LC, P) automatically adopt proper core policy and adjust cores to respond to the changing needs. And for T1(LC, C), core policies are changed about 20 times in a period of 600 seconds. Therefore, for QWin, not only target SLOs of LC tenants can be guaranteed, but also the bandwidth of BE tenants is maximized.

\section{Related Work}
\textbf{Dynamic Core allocation.} When deciding how to allocate threads or cores for applications, previous works have allocated cores by monitoring performance metrics, utilization, or statistics~\cite{heracles15,arachne18,shenango19,caladan20,perfiso18,parties19}.
Arachne~\cite{arachne18} uses load factor and utilization of allocated cores to increase or decrease cores for LC tenants. Both Caladan~\cite{caladan20} and Shenango~\cite{shenango19} rely on queueing delay to adjust core allocation for LC tenants. PerfIso~\cite{perfiso18} ensures that there are always servals idle cores available for LC tenants to accommodate bursty load.
Both Heracles~\cite{heracles15} and Parties~\cite{parties19} adjust cores allocation between LC and BE tenants depending on each tenant’s slack (the difference between the target SLO and the measured tail latency). 
However, all these approaches allocate cores at a coarse granularity, and they do not calculate the accurate number of cores according to target SLO. 
And core adjustment is incremental, and latency suffers during the convergence of adjustment.
Moreover, they do not consider the underlying storage device which could have long tail latency and large service time fluctuations. 

\textbf{Request Scheduling.} Several systems~\cite{cake12,reflex17,pm14,sncm16,wc17} adopt a sharing model and combine priority-based scheduling with rate limiting control to guarantee target SLOs for LC tenants.
Cake~\cite{cake12} adjusts proportional share and reservation of cores between an LC tenant and a BE tenant. Reflex~\cite{reflex17} first uses target SLOs and offline profiled data of NVMe SSD to determine the upper rate limits, and schedules LC and BE requests based on priorities and rate limiting.
PriorityMeister\cite{pm14}, SNC-Meister\cite{sncm16} and WorkloadCompactor\cite{wc17} determine each tenant's priority and rate limits through offline analysis of traces using mathematical methods (e.g. DNC\cite{dnc01} and SNC\cite{snc15}), or through target SLO and offline profiled data of NVMe SSD.
Some Cores are still shared between LC and BE tenants in these systems, easily leading target SLO violations.
Besides, the above proactive approaches would cause SLO violations or low utilization if the real access patterns of requests deviate from the workload traces. 
Several scheduling optimizations have been proposed to reduce tail latency (e.g. Rein\cite{rein17}, Few-to-Many\cite{fm15}, Tail Control\cite{tailcontrol16}, Shinjuku\cite{shinjuku19}, and Minos\cite{minos19}). These works mainly focus on different size of requests within an application. They consider reducing tail latency for key-value stores\cite{rein17,shinjuku19,minos19}, or require application that are dynamically parallelizable\cite{fm15,tailcontrol16}. However, guarantee tail latency SLO is beyond the scope of these works.

\textbf{Replication-based approaches.} Existing works focus on reducing tail latency by making use of duplicate requests (e.g. MittOS\cite{mittos17}, and CosTLO\cite{costlo15}) and adaptive replica selection (e.g. C3\cite{rs2} and NetRS\cite{NetRS18}). These approaches consider capability of all servers, and select a faster server to serve requests.
QWin is orthogonal to those work as it guarantees target SLO within a storage server.
We are interested in exploring ways of integrating these techniques with QWin in the future.

\section{Conclusion}
This paper presents QWin, a tail latency SLO aware core allocation that enforces target SLO for LC tenants and enhances bandwidth of BE tenants.
The effectiveness of QWin comes from its key ideas: 1) an SLO-to-core calculation model, which accurately calculates the number of cores, making the allocation of core satisfied at a time without gradually converging. 2) a flexible request-based window, which quantifies the definitive runtime load for the SLO-to-core calculation model; 3) three core policies, which determine the frequency for core adjustment; 4) an autonomous core allocation, which adjusts cores without any dedicated core.
These contributions allow QWin to guarantee target SLO of LC tenants meanwhile increase bandwidth of BE tenants by up to 31x.

%-------------------------------------------------------------------------------
\section*{Acknowledgments}
We thank our shepherd Zhiwei Xu and Xinwu Liu for thier useful feedback. This work is supported by the National Key Research and Development Program of China (2016YFB1000202) and Alibaba Innovative Research (No.11969786).
%-------------------------------------------------------------------------------

\bibliographystyle{unsrt}
\bibliography{references}

%%%%%%%%%%%%%%%%%%%%%%%%%%%%%%%%%%%%%%%%%%%%%%%%%%%%%%%%%%%%%%%%%%%%%%%%%%%%%%%%
\end{document}